\documentstyle[amssymb,prl,aps,epsf,floats]{revtex}

\begin{document}
\draft

\title{Analytic solution for a class of turbulence problems }
\author{M. Vlad$^1$, F.\ Spineanu$^1$, J. H. Misguich$^2$, and R.\ Balescu$^3$}
\address{$^1$National Institute for Laser, Plasma and Radiation Physics, \\
Association Euratom-NASTI Romania, P.O.Box MG-36, Magurele, Bucharest, Romania\\
$^2$Association Euratom-C.E.A. sur la Fusion, C.E.A.-Cadarache, \\
F-13108 Saint-Paul-lez-Durance, France\\
$^3$Association Euratom-Etat Belge sur la Fusion,\\
Universit\'{e} Libre de Bruxelles, Bd.du Triomphe, 1050 Bruxelles, Belgium\\
}
\maketitle

\begin{abstract}
An exact analytical method for determining the Lagrangian velocity
correlation and the diffusion coefficient for particles moving in\ a
stochastic velocity field is derived. It applies to \ divergence-free
2-dimensional Gaussian stochastic fields which are stationary, homogeneous
and have factorized Eulerian correlations.
\end{abstract}

%


Test particle motion in stochastic velocity fields is a generic problem in
various topics of fluid and plasma turbulence or solid state physics \cite
{BG}. The main difficulty in determining the resulting time dependent
(running) diffusion coefficients and mean square displacements consists in
calculating the Lagrangian velocity correlation function (LVC). This is a
very complex quantity which requires the knowledge of the statistical
properties of the stochastic trajectories determined by the random velocity
field. The vast majority of existing works employ under various guises the
Corrsin approximation \cite{mccomb}. The physical parameter which
characterizes such process of diffusion by continuous movements is the Kubo
number $K$ (defined below) which measures particle's capacity of exploring
the space structure of the stochastic velocity field before the latter
changes. In the weak turbulence case $K\ll 1,$ particle motion is of
Brownian type and the results are well established. In the strong turbulence
case ($K>1),$ this structure of the velocity field has an important
influence on the LVC and on the scaling of the diffusion coefficient in $K$.
This influence is most effective in the special case of \ 2-dimensional
divergence-free stochastic velocity fields and consists in a dynamical
trapping of the trajectories in the structure of the field. The existing
analytical methods completely fail in describing this process \cite
{kraichnan} and the studies usually rely on direct numerical simulations of
particle trajectories \cite{jacques}, on asymptotic methods such as the
renormalization group techniques \cite{BG} or on qualitative estimates \cite
{isichenko}. In recent works \cite{VSMB1} a rather different statistical
approach (the decorrelation trajectory method) was proposed for determining
the LVC for given Eulerian correlation of the velocity field. The case of collisional particles was treated in \cite{VSMB2}. We prove here
that, in the special case of 2-dimensional divergence-free velocity fields,
under the assumptions mentioned below, this method yields the exact
analytical expression of the Lagrangian velocity correlation valid for
arbitrary value of the Kubo number. The assumptions concern the statistical
properties of the stochastic velocity field and are rather natural for a
large class of physical processes. It is considered to be a stationary and
homogeneous Gaussian (normal) stochastic field, either static or time
dependent with statistically independent space and time variations such that
the Eulerian correlation function has a factorized structure as in Eq.(\ref
{pec}) below.

\bigskip
Particle motion in a 2-dimensional stochastic velocity field is described by
the nonlinear Langevin equation: 
\begin{equation}
\frac{d{\bf x}(t)}{dt}={\bf v}({\bf x}(t),t),\qquad {\bf x}(0)={\bf 0}
\label{1}
\end{equation}
where ${\bf x}(t)$ represents the trajectory in Cartesian coordinates ${\bf x%
}\equiv (x_{1},x_{2}).$ The stochastic velocity field ${\bf v}({\bf x},t)$
is \ divergence-free: ${\bf \nabla \cdot v}({\bf x},t)=0$ and thus its two
components $v_{1}$ an $v_{2}$ can be determined from a stochastic scalar
field, the stream function (or potential) $\phi ({\bf x},t)$, as: 
\begin{equation}
{\bf v}({\bf x},t)={\bf \nabla }\times \phi ({\bf x},t){\bf e}_{z}=\left( 
\frac{\partial }{\partial x_{2}},\quad -\frac{\partial }{\partial x_{1}}%
\right) \phi ({\bf x},t)  \label{vd}
\end{equation}
where ${\bf e}_{z}$ is the versor along the $z$ axis. The stochastic stream
function $\phi ({\bf x},t)$ is considered to be Gaussian, stationary and
homogeneous. Since the velocity components are derivatives of $\phi ({\bf x}%
,t)$, they are Gaussian, stationary and homogeneous as well. We assume that
they have zero averages: 
\begin{equation}
\left\langle \phi ({\bf x},t)\right\rangle =0,\quad \left\langle {\bf v}(%
{\bf x},t)\right\rangle ={\bf 0.}  \label{av0}
\end{equation}
The Eulerian two-point correlation function (EC) of $\phi ({\bf x},t)$ is
assumed to be of the form: 
\begin{equation}
E({\bf x},t)\equiv \left\langle \phi ({\bf 0},0)\,\phi ({\bf x}%
,t)\right\rangle =\beta ^{2}{\cal E}({\bf x})\,h(t)  \label{pec}
\end{equation}
where $\beta $ measures{\em \ } the amplitude of the stream function
fluctuations and $\left\langle {}\right\rangle $ denotes the statistical
average over the realizations of $\phi ({\bf x},t).$ ${\cal E}({\bf x)}$ is
a dimensionless function having a maximum at ${\bf x}={\bf 0}${\bf , }where
its value is ${\cal E}{\bf (0})=1$, and which tends to zero as $\left| {\bf x%
}\right| \rightarrow \infty $.\ It actually depends on the dimensionless
variable ${\bf x}/\lambda $, where $\lambda $ is the correlation length. $%
h(t)$ is a dimensionless, decreasing function of time varying from $h(0)=1$
to $h(\infty )=0$. It depends on the dimensionless ratio $t/\tau _{c}$,
where $\tau _{c}$ is the correlation time. The Kubo number is defined as the
ratio of the average distance covered by the particles during $\tau _{c}$ to 
$\lambda $ : $K=V\,\tau _{c}/\lambda $ \thinspace where $V=\beta /\lambda $
measures the amplitude of the fluctuating velocity. Using the definition (%
\ref{vd}) of the velocity, the two-point Eulerian correlations of the
velocity components and \ the potential-velocity correlations{\em \ } are
obtained from $E({\bf x},t)$ as: 
\begin{eqnarray}
E_{11} &=&-\frac{\partial ^{2}}{\partial x_{2}^{2}}E,\quad E_{22}=-\frac{%
\partial ^{2}}{\partial x_{1}^{2}}E,\quad E_{12}=\frac{\partial ^{2}}{%
\partial x_{1}\partial x_{2}}E,  \label{cev} \\
E_{1\phi } &=&-E_{\phi 1}=-\frac{\partial }{\partial x_{2}}E,\quad E_{2\phi
}=-E_{\phi 2}=\frac{\partial }{\partial x_{1}}E  \nonumber
\end{eqnarray}
where $E_{ij}({\bf x},t)\equiv \left\langle v_{i}({\bf 0},0)\,v_{j}({\bf x}%
,t)\right\rangle $ \thinspace and $E_{\phi i}\equiv \left\langle \phi ({\bf 0%
},0)\,\,v_{i}({\bf x},t)\right\rangle .$

\smallskip Starting from this statistical description of the stochastic
stream function, we will determine the Lagrangian velocity correlation
(LVC),\ defined by:

\begin{equation}
L_{ij}(t)\equiv \left\langle v_i({\bf x}(0),0)\,v_j({\bf x}%
(t),t)\right\rangle .  \label{CL}
\end{equation}
The mean square displacement and the running diffusion coefficient are
determined by this function: 
\begin{equation}
\left\langle x_i^2(t)\right\rangle =2\int_0^td\tau \;L_{ii}(\tau )\;(t-\tau
),  \label{MSD}
\end{equation}
\begin{equation}
D_i(t)=\int_0^td\tau \;L_{ii}(\tau ).  \label{D}
\end{equation}

For small Kubo numbers (quasilinear regime), the results are well
established: the diffusion coefficient is $D_{QL}=(\lambda ^{2}/\tau
_{c})K^{2}.$ At large $K$ the time variation of the stochastic potential is
slow and the trajectories can follow approximately the contour lines of $%
\phi ({\bf x},t).$ This produces a trapping effect : the trajectories are
confined for long periods in small regions.\ A typical trajectory shows an
alternation of large displacements and trapping events. The latter appear
when the particles are close to the maxima or minima of the stream function
and consists of trajectory winding on almost closed small size paths. The
large displacements are produced when the trajectories are at small absolute
values of the stream function.\ 

\bigskip
The main idea in our method is to study the Langevin equation (\ref{1}) in
subensembles (S) of the realizations of the stochastic field which are
determined by given values of the stream function and of the velocity in the
starting point of the trajectories: 
\begin{equation}
\phi ({\bf 0},0)=\phi ^{0},\quad {\bf v}({\bf 0},0)={\bf v}^{0}.  \label{2}
\end{equation}
The LVC for the whole set of realizations can is obtained by summing up the
contributions of all subensembles (\ref{2}): 
\begin{equation}
L_{ij}(t)=\int \int d\phi ^{0}\,d{\bf v}^{0}\,P_{1}(\phi ^{0})\,P_{1}({\bf v}%
^{0})\,L_{ij}^{s}(t)  \label{3}
\end{equation}
where $L_{ij}^{s}(t)$ is the LVC in (S) and $\,P_{1}(\phi ^{0})\,$, $P_{1}(%
{\bf v}^{0})\,$ are the Gaussian (normal) probability densities for the
initial stream function and respectively for the initial velocity. As shown
below, there are two important advantages determined by this procedure: (i)
the LVC can be determined from one-point subensemble averages and (ii) the
invariance of the stream function along the trajectory can be used for
obtaining the average of the Lagrangian velocity and the LVC in (S).

The stream function and the velocity reduced in the subensemble (S) are
still Gaussian stochastic fields but non-stationary and non-homogeneous and
with space-time dependent average values: 
\begin{equation}
\Phi ^{S}({\bf x},t)\equiv \left\langle \phi ({\bf x},t)\right\rangle
_{S}=\phi ^{0}\frac{E({\bf x},t)}{\beta ^{2}}+v_{j}^{0}\frac{E_{j\phi }({\bf %
x},t)}{V^{2}},  \label{fim}
\end{equation}
\begin{equation}
V_{i}^{S}({\bf x},t)\equiv \left\langle v_{i}({\bf x},t)\right\rangle
_{S}=\phi ^{0}\frac{E_{\phi i}({\bf x},t)}{\beta ^{2}}+v_{j}^{0}\frac{E_{ji}(%
{\bf x},t)}{V^{2}}  \label{vim}
\end{equation}
where $\left\langle {}\right\rangle _{S}$ represents the average over the
realizations in the subensemble (S). These averages are determined using
conditional probability distribution; they are equal in ${\bf x}={\bf 0}$
and $t=0$ to the parameters determining (S): $\Phi ^{S}({\bf 0},0)=\phi
^{0},\;V_{i}^{S}({\bf 0},0)=v_{i}^{0}$ and decay to zero as ${\bf x}%
\rightarrow \infty $ and/or $t\rightarrow \infty $. A relation similar to (%
\ref{vd}) can easily be deduced: 
\begin{equation}
{\bf V}^{S}({\bf x},t)=\left( \frac{\partial }{\partial x_{2}},\quad -\frac{%
\partial }{\partial x_{1}}\right) \Phi ^{S}({\bf x},t)  \label{vfi}
\end{equation}
which shows that the average velocity in the subensemble (S) is \
divergence-free: ${\bf \nabla \cdot V}^{S}({\bf x},t)=0.$ We have thus
identified in the zero-average{\em \ } stochastic velocity field {\it a set
of average velocities}{\bf \ }(labeled by $\phi ^{0},$ ${\bf v}^{0})$ which
contain the statistical characteristics of the velocity field (the
correlation and the constraint imposed in the problem, i.e. the zero
divergence condition). The LVC\ in (S) is: 
\begin{equation}
L_{ij}^{S}(t)\equiv \left\langle v_{i}({\bf 0},0)\,v_{j}({\bf x(t)}%
,t)\right\rangle _{S}=v_{i}^{0}\left\langle v_{j}({\bf x(t)},t)\right\rangle
_{S}  \label{lvs}
\end{equation}
and thus the problem reduces to the determination of the average Lagrangian
velocity in each subensemble.

\bigskip
We consider first the static case $\phi ({\bf x})$ ($\tau _{c}\rightarrow
\infty ,$ $K\rightarrow \infty ,$ and the EC of the stream function $E({\bf x%
})$ is independent of time). The stream function is an invariant of the
motion ($\phi ({\bf x}(t))=\phi ^{0}$ in each realization in (S)) and: 
\begin{equation}
\left\langle \phi ({\bf x}(t))\right\rangle _{S}=\phi ^{0}  \label{fims}
\end{equation}
at any time. A {\it deterministic trajectory} ${\bf X}(t;S)$ can be defined
in each subensemble (S) such that the average of the Eulerian stream
function in (S) (Eq.(\ref{fim})) calculated along this trajectory equals the
average Lagrangian stream function (\ref{fims}): 
\begin{equation}
\Phi ^{S}({\bf X}(t;S))=\left\langle \phi ({\bf x}(t))\right\rangle
_{S}=\phi ^{0}.  \label{e2}
\end{equation}
Since the Eulerian average potential (\ref{fim}) has the value $\phi ^{0}$
in ${\bf x=0}$, the trajectory ${\bf X}(t;S)$ can be determined from the
following Hamiltonian system of equations with $\Phi ^{S}({\bf X})$ as
Hamiltonian function: 
\begin{equation}
\frac{d{\bf X}(t;S)}{dt}=\left( \frac{\partial }{\partial X_{2}},\quad -%
\frac{\partial }{\partial X_{1}}\right) \Phi ^{S}({\bf X}(t;S))
\label{dectr}
\end{equation}
and with the initial condition ${\bf X(}0;S)={\bf 0.}$ The trajectory in
each realization in (S) can be referred to this deterministic
(realization-independent) trajectory, ${\bf x}(t)={\bf X}(t;S)+\delta {\bf x}%
(t).$ Using the definition of the velocity (\ref{vd}) and expressing the
space derivatives as derivatives with respect to the deterministic part of
the trajectory in each realization and then averaging in the subensemble
(S), the average Lagrangian velocity in (S) is obtained as: 
\begin{equation}
\left\langle {\bf v}({\bf x}(t))\right\rangle _{S}={\bf V}^{S}\left( {\bf X}%
(t;S)\right)  \label{alvs}
\end{equation}
where Eq.(\ref{vfi}) was used. Thus, the average Lagrangian velocity in the
subensemble (S) is just the corresponding Eulerian quantity calculated along
the deterministic trajectory ${\bf X}(t;S).$ Since the latter is determined
by solving Eq.(\ref{dectr}), where the r.h.s. is the average Lagrangian
velocity, it follows that{\em \ }${\bf X}(t;S)$ is \ precisely the average
trajectory in (S) : 
\begin{equation}
{\bf X}(t;S)=\left\langle {\bf x}(t)\right\rangle _{S}.  \label{trm}
\end{equation}

\bigskip
Similar results are obtained in the time-dependent case $\phi ({\bf x},t)$
(finite $\tau _{c}$ and $K)$ if the space and time dependences are
statistically independent in the sense that the EC of $\phi ({\bf x},t)$ is
given by Eq.(\ref{pec}). The stream function is not a true invariant of the
motion. However the velocity is still perpendicular to ${\bf \nabla }\phi (%
{\bf x}(t),t)$ at any moment and only the explicit time-dependence
contributes to the variation of $\phi $ along the trajectory: 
\begin{equation}
\frac{d\phi ({\bf x}(t),t)}{dt}=\frac{\partial \phi ({\bf x}(t),t)}{\partial
t}.  \label{defi}
\end{equation}
Due to the factorized EC (\ref{pec}) considered here, the average Eulerian
stream function and velocity (\ref{fim}), (\ref{vim}) can be written as: 
\begin{equation}
\Phi ^{S}({\bf x},t)=\Phi ^{S}({\bf x})\,h(t),  \label{tdfi}
\end{equation}
\begin{equation}
{\bf V}^{S}({\bf x},t)={\bf V}^{S}({\bf x})\,h(t)  \label{tdv}
\end{equation}
where $\Phi ^{S}({\bf x})$ and ${\bf V}^{S}({\bf x})$ are the corresponding
quantities in the static case. We define in (S) a deterministic trajectory $%
\overline{{\bf X}}(t;S)$ as the solution of the time dependent Hamiltonian
system with $\Phi ^{S}(\overline{{\bf X}},t)$ as Hamiltonian function.
Performing the change of variable $t\rightarrow \theta (t)=\int_{0}^{t}d\tau
\,h(\tau ),$ the time dependent Hamiltonian system reduces to Eq.(\ref{dectr}%
) and thus the trajectory $\overline{{\bf X}}(t;S)$ can be written as: 
\begin{equation}
\overline{{\bf X}}(t;S)={\bf X}(\theta (t);S)  \label{trmtd}
\end{equation}
where ${\bf X}(\theta ;S)$ is the deterministic trajectory obtained in the
static case. On the other hand, the average Lagrangian potential
corresponding to Eq.(\ref{tdfi}) can be written as $\left\langle \phi ({\bf x%
}(t),t)\right\rangle _{S}=G(t)\,h(t)$ where the factor $h(t)$ ''propagates''
unchanged from the Eulerian average to the Lagrangian one, and $G(t)$ is the
contribution of the space dependence. Taking the time derivative of this
equation one obtains using Eq.(\ref{defi}): 
\[
\frac{d}{dt}\left\langle \phi ({\bf x}(t),t)\right\rangle _{S}=%
\mathrel{\mathop{\lim }\limits_{\delta t\rightarrow 0}}%
\frac{\left\langle \phi ({\bf x}(t),t+\delta t)\right\rangle
_{S}-\left\langle \phi ({\bf x}(t),t)\right\rangle _{S}}{\delta t}. 
\]
It follows that $d\left\langle \phi ({\bf x}(t),t)\right\rangle
_{S}/dt=G(t)dh/dt$ and thus $G(t)=\phi ^{0}$ and the average Lagrangian
stream function in (S) is: 
\begin{equation}
\left\langle \phi ({\bf x}(t),t)\right\rangle _{S}=\phi ^{0}\,h(t).
\label{fimst}
\end{equation}
Using Eq.(\ref{fimst}) and (\ref{tdfi}) and the definition of ${\bf \ }$ $%
\overline{{\bf X}}(t;S)$, one finds that the average Eulerian stream
function calculated along the deterministic trajectory ${\bf \ }\overline{%
{\bf X}}(t;S)$ equals, as in the static case, the average Lagrangian stream
function: 
\begin{equation}
\Phi ^{S}\left( \overline{{\bf X}}(t;S)\right) \,h(t)=\left\langle \phi (%
{\bf x}(t),t)\right\rangle _{S}=\phi ^{0}\,h(t)  \label{e2t}
\end{equation}
Following the same arguments as in the static case, the average Lagrangian
velocity in (S) is determined as: 
\begin{equation}
\left\langle {\bf v}({\bf x}(t),t)\right\rangle _{S}={\bf V}^{S}\left( 
\overline{{\bf X}}(t;S)\right) \,h(t)  \label{alvt}
\end{equation}
and the deterministic trajectory $\overline{{\bf X}}(t;S)$ is the average
trajectory in (S): 
\begin{equation}
\overline{{\bf X}}(t;S)=\left\langle {\bf x}(t)\right\rangle _{S}.
\label{trmt}
\end{equation}

\bigskip
We have thus obtained the subensemble averages of the Lagrangian stream
function and velocity. These averages of random functions of random
arguments appear to be equal to the average functions evaluated at the
average argument. We note that this surprisingly simple result which is
usually wrong for stochastic functions is exact for the special case studied
here. This property is essentially due to the very strong constraint imposed
by the invariance of the stream function along the trajectories.

\bigskip
The LVC (\ref{CL}) and the running diffusion coefficient (\ref{D}), are
determined using Eqs. (\ref{3}) and (\ref{lvs}) as: 
\begin{equation}
L_{ij}(t)=\delta _{ij}V^{2}\,F^{\prime }\left( K\theta (t)\right) \,h(t),
\label{L}
\end{equation}
\begin{equation}
D(t)=\frac{\lambda ^{2}}{\tau _{c}}\,K\,F\left( K\theta (t)\right)
\label{dt}
\end{equation}
where $F^{\prime }(\tau )$ is the derivative of the function $F(\tau )$
which is defined by: 
\begin{equation}
F(\tau )\equiv \frac{1}{\sqrt{2\pi }}\int \int_{0}^{\infty }dpdu\,u^{3}\exp
\left( -\frac{u^{2}(1+p^{2})}{2}\right) X(\tau ;u,p).  \label{F}
\end{equation}
$X(\tau ;u,p)$ is the component of the average trajectory (normalized by $%
\lambda )$ along ${\bf v}^{0}$ and the dimensionless parameters $u\equiv
\left| {\bf v}^{0}\right| /V,$ $p\equiv \phi ^{0}/u\beta $ determine the
subensemble (S). It is the solution of Eq.(\ref{dectr}). The function $%
\theta $ in Eqs.(\ref{L}), (\ref{dt}) is defined by: 
\begin{equation}
\theta (t)=\int_{0}^{t}h(t_{1})\,dt_{1}.  \label{teta}
\end{equation}

We note that Eqs.(\ref{L})-(\ref{teta}) represent the exact solution for the
diffusion problem studied here (both for static and time-dependent case).
Two functions are involved in the expressions for the LVC and $D(t).$ One is
the time dependence $h(t)$ of the EC of the stream function. This accounts
for the explicit time decorrelation and remains unchanged when passing from
Eulerian to Lagrangian quantities. The other is the function $F(\tau )$
which results from the space dependence of the EC of the stream function
(i.e. from the Lagrangian nonlinearity). It is obtained as an integral over
the subensembles (S) of the average displacements along the initial velocity 
${\bf v}^{0}.$

\begin{figure}[t]
\centerline{\epsfxsize=14cm\epsfbox{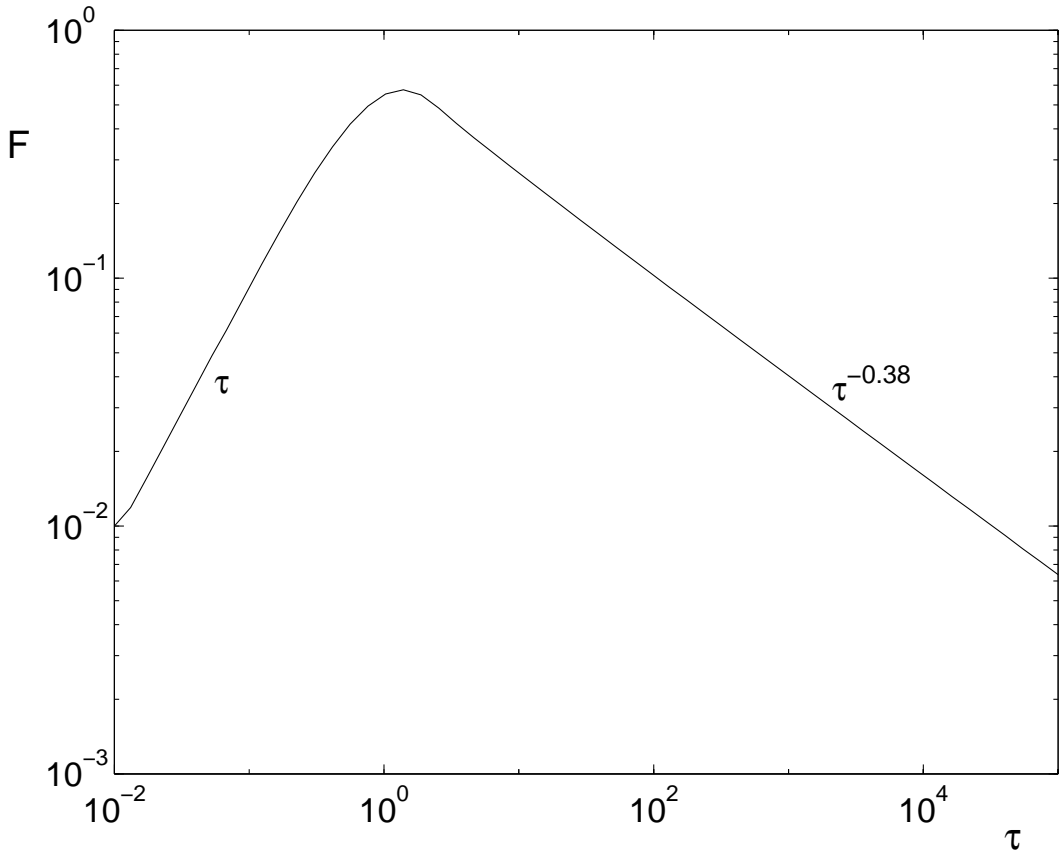}}
\caption{The function $F(\protect\tau )$ defined in Eq.(\ref{F})}
\end{figure}

\bigskip
The trajectories obtained from Eq.(\ref{dectr}) lie on closed paths (except
for $\phi ^{0}=0$ which correspond to a straight line along ${\bf v}^{0}$).\
The size of these paths is large for small $\left| \phi ^{0}\right| $ and
it decreases as $\left| \phi ^{0}\right| $ increases. At small time $\tau ,$ 
$X(\tau ;u,p)\cong u\tau $ and $F(\tau )\cong \tau .$ At large time $\tau ,$
the trajectory ${\bf X}(\tau ;u,p)$ turns periodically along the
corresponding path.\ The period grows with the size of the path. Thus, for a
given time $\tau ,$ the trajectories corresponding to small paths (large $%
\left| \phi ^{0}\right| )$ rotate many times while those along large enough
paths (small $\left| \phi ^{0}\right| )$ are still opened.\ Consequently,
when calculating the integrals in Eq.(\ref{F}), the contribution of the
small paths (large $\left| \phi ^{0}\right| )$ \ is progressively eliminated
as $\tau $ increases due to incoherent mixing. As $\tau $ increases, smaller
and smaller intervals of $\phi ^{0}$ around $\phi ^{0}=0$ effectively
contribute to the function $F(\tau )$ which consequently decays to zero as $%
\tau \rightarrow \infty .$ Thus, the function $F$ accounts for the dynamical
trapping of the trajectories. The characteristic features of this
self-consistent trapping process observed in the numerical simulations are
recovered in the structure of this function. It shows that only a part of
the trajectories (which are not yet trapped) effectively contribute to the
value of the diffusion coefficient at that moment. These are the particles
that move on the large size contour lines of the stochastic stream function
which correspond to $\phi \cong 0.$ The trapping process is evidenced at
large time and determines the decay of $F$ as $F(\tau )\sim \tau ^{-\alpha }.
$ The function $F(\tau )$ obtained for ${\cal E}({\bf x})=1/(1+{\bf x}%
^{2}/2\lambda ^{2})$ is presented in Fig.1 where the two regimes are
observed. The value of $\alpha $ for this case is $\alpha =0.38.$ The
modification of ${\cal E}({\bf x})$ determines a variation of $\alpha $
around this value.

\bigskip
In a static stream function ($K,\tau _{c}\rightarrow \infty ,$ $h(t)=1)$ the
particles move along the ''frozen'' contour lines of $\phi ({\bf x})$ and
the process is subdiffusive. This can easily be recovered in the general
result (\ref{L})-(\ref{teta}).\ The average trajectories in the subensembles
(S) are periodic functions of time and the diffusion coefficient (\ref{dt})
is $D(t)=\beta \,F\left( Vt/\lambda \right) .$ It goes to zero when $%
t\rightarrow \infty $ as $D(t)=\beta \left( Vt/\lambda \right) ^{-\alpha }$
and the mean square displacement is $\left\langle x^{2}(t)\right\rangle \sim
t^{1-\alpha }.$

\bigskip
In the time dependent case, the time variation of the stream function
determines a decorrelation effect which leads to a diffusive process. This
is reflected in the average trajectories in the subensembles (S) (determined
by Eqs.(\ref{trmt}), (\ref{trmtd})) which are not anymore periodic functions
of time but all of them saturate as $t\rightarrow \infty $ (possibly after
performing many rotations around the corresponding paths). Consequently, the
decay of the function $F$ saturates at $F(K\,\theta _{\infty })$ where $%
\theta _{\infty }$ is a constant of order 1 obtained as the limit $%
t\rightarrow \infty $ of Eq.(\ref{teta}) and the asymptotic diffusion
coefficient is: 
\begin{equation}
D=\frac{\lambda ^{2}}{\tau _{c}}K\,F\left( K\,\theta _{\infty }\right) .
\label{das}
\end{equation}
In the limit of small $K,$ the quasilinear result is recovered from Eq.(\ref
{das}) and at large $K$ when trapping is important $D\approx (\lambda
^{2}/\tau _{c})K^{1-\alpha }.$

\bigskip
In conclusion, we have presented here an exact solution for the turbulent
diffusion problem for a class of velocity fields. We have obtained
analytical expressions for the LVC and $D(t)$ which are valid for arbitrary
values of the Kubo number and describes the complicated process of dynamic
trajectory trapping in the structure of the stochastic field. The basic idea
of the method consists of determining the LVC by means of a set of average
Lagrangian velocities estimated in subensembles of realizations of the
stochastic field which are defined taking into account the invariants of the
motion.\ It can be extended, at least as a new type of approximation, to
other types of stochastic velocity fields.

\bigskip
This work has benefited of the NATO Linkage Grant CRG.LG 971484 which is
acknowledged.


\begin{references}
\bibitem{BG}  J. P. Bouchaud and A. George, Phys. Reports {\bf 195, }128
(1990).

\bibitem{mccomb}  W.\ D.\ McComb, The Physics of Fluid Turbulence
(Clarendon, Oxford,\ 1990).

\bibitem{kraichnan}  R. H. Kraichnan, Phys. Fluids {\bf 19}, 22 (1970).

\bibitem{jacques}  J.-D. Reuss and J.\ H.\ Misguich, Phys. Rev. E {\bf 54},
1857 (1996); J.-D. Reuss, M.\ Vlad and J.\ H.\ Misguich, Phys. Lett. A {\bf %
241}, 94 (1998).

\bibitem{isichenko}  M.\ B.\ Isichenko, Plasma Phys. Contr.\ Fusion {\bf 33}%
, 809 (1991).

\bibitem{VSMB1}  M.\ Vlad, F.\ Spineanu, J. H.\ Misguich and R.\ Balescu,
Phys. Rev. E {\bf 58}, 7359 (1998).

\bibitem{VSMB2}  M.\ Vlad, F.\ Spineanu, J. H.\ Misguich and R.\ Balescu,
Phys. Rev. E {\bf 61}, 3023 (2000).
\end{references}
\end{document}